\documentclass[sigconf, screen]{acmart}
\usepackage{threeparttable}

\AtBeginDocument{%
  }

\setcopyright{none}

\acmConference[arxiv 2026]{}{June 5, 2026}{World, Earth}

\begin{document}

\title{ANNS-AMP: Accelerating Approximate Nearest Neighbor Search via Adaptive Mixed-Precision Computing}
\author{Mingkai Chen, Cheng Liu, Shengwen Liang, Lei Zhang, Xiaowei Li, and Huawei Li}
\affiliation{%
  \institution{Institute of Computing Technology, Chinese Academy of Sciences and University of Chinese Academy of Sciences}
  \city{Beijing}
  \country{China}}
  



\begin{abstract}

Approximate nearest neighbor search (ANNS) is a critical kernel in modern applications such as large language model (LLM) inference and recommendation systems. However, its efficiency is fundamentally limited by the need to compute distances between a query and a massive number of high-dimensional vectors, most of which are non-neighbors. Existing approaches reduce redundancy via index optimization or early termination, but remain constrained by fixed-precision computation, leading to unnecessary arithmetic and memory bandwidth overhead. 
This paper presents ANNS-AMP, an adaptive mixed-precision framework and accelerator that adapts the precision of distance computation to the characteristics of queries and data distribution. The key insight is that different regions of the vector space require different levels of precision to preserve top-\emph{k} accuracy. ANNS-AMP leverages the clustered structure of product-quantization (PQ)-based indices and introduces a lightweight predictor to determine cluster-level precision at runtime based on features such as scale, radius, and query distance. To efficiently realize variable-precision execution, we design a bit-serial accelerator with a bit-interleaved data layout, enabling throughput to scale with reduced precision while mitigating memory bandwidth bottlenecks and load imbalance through a greedy scheduling strategy. Moreover, the runtime predictor can also reuse the bit-serial computing array for efficient runtime prediction and can be fitted to the ANNS pipeline without performance penalty. According to our experiments on representative datasets, ANNS-AMP achieves 163.76×, 10.57×, and 2.06× performance speedups on average, and reduces average energy consumption by 1100.00×, 39.41×, and 6.66× compared to CPU, GPU, and customized ANNS accelerator baselines, respectively, while maintaining accuracy loss below 2.7\%. These results demonstrate that adaptive mixed-precision computing is a promising direction for efficient large-scale ANNS.

\end{abstract}


\keywords{Approximate Nearest Neighbor Search, Mixed Precision Computing, Bit-Serial Architecture} 

\maketitle

\section{Introduction}

Approximate nearest neighbor search (ANNS) is a fundamental building block in many modern applications, including large language model (LLM)-based generation and recommendation systems. In these scenarios, data from diverse modalities are embedded into high-dimensional vector spaces, where semantic similarity is reflected by proximity. However, the size of the base vector corpus is often orders of magnitude larger than the set of true nearest neighbors \cite{refp24}, resulting in substantial computational and memory overhead for distance evaluation over non-relevant vectors. This imbalance motivates the need for efficient algorithmic and architectural techniques to reduce redundant computations while maintaining accuracy.

Prior work has explored various strategies to mitigate redundancy in ANNS~\cite{refp01, refp02, refp03, refp04, refp05, refp06, refp07, refp08}. A line of research focuses on optimizing index construction. For example, FANNS~\cite{refp06}, VDTuner~\cite{refp02}, and DRIM-ANN~\cite{refp09} tune index parameters offline to improve search efficiency under accuracy constraints. However, once constructed, the index remains fixed during query processing, limiting adaptability to varying query distributions. Other approaches exploit early termination mechanisms to prune redundant computations at different granularities, including vector-level~\cite{refp03}, dimension-level~\cite{refp45}, and bit-level~\cite{refp24}. These methods rely on dynamically updated top-\emph{k} thresholds, which are often loose in early stages of the search and thus limit pruning effectiveness. Moreover, in cluster-based ANNS methods, lookup-table (LUT) construction requires computing and storing distances for all entries, where early termination cannot be directly applied. JUNO~\cite{refp05} partially addresses this issue by pruning LUT construction on GPUs using regression-based thresholds over grid-partitioned subspaces. However, grid-based partitioning in low-dimensional subspaces does not generalize well to high-dimensional vector spaces, where neighbors may not be locally clustered in each subspace~\cite{refp14}. As a result, aggressive pruning may degrade accuracy, especially for high-dimensional data.

Instead of uniformly computing distances at full precision, we seek to provide just-enough precision for each computation under accuracy constraints, thereby reducing both arithmetic cost and memory bandwidth. Importantly, the required precision depends on both the query and the distribution of the centroids of clusters, and thus varies across queries and regions of the search space. To enable this, we develop a lightweight precision predictor that dynamically determines the appropriate computing precision. Leveraging the clustered structure of typical PQ-based ANNS, we perform prediction at the cluster level rather than for individual vectors, significantly reducing overhead. This design enables efficient runtime adaptation of precision across different regions of the vector space. To fully exploit this opportunity, we further adopt a bit-serial computing paradigm that naturally supports mixed-precision execution, allowing throughput to scale with reduced precision. Compared to conventional CPUs and GPUs, which offer limited support for sub-8-bit arithmetic, this approach unlocks finer-grained performance–efficiency trade-offs.

Despite its promise, mixed-precision acceleration for ANNS introduces several challenges. First, within the clustered index structure of typical PQ-based ANNS, partitioning high-dimensional vector spaces into fine-grained subspaces for precision control can be costly and ineffective when using conventional grid-based methods, which do not scale well with dimensionality. Second, accurately predicting precision requires identifying representative features that capture the relationship between query distribution and distance sensitivity. Third, translating reduced precision into actual performance gains necessitates hardware support that can efficiently handle variable-precision computation and memory access without introducing load imbalance or bandwidth bottlenecks.

To address these challenges, we propose ANNS-AMP, an adaptive mixed-precision framework and accelerator for efficient ANNS. We begin by partitioning the vector space of the clustered ANNS structure. To adapt to high-dimensional data distributions, we introduce dimension-wise splitting prior to clustering, with adjustable granularity. Based on an analysis of the top-\emph{k} selection process, we extract key features for each cluster, including scale, radius, and distance to the query, and train a regression model offline to predict precision requirements at runtime. With the predictor, the distance calculation is adaptively performed at varying precisions for each query and cluster accordingly. The resulting mixed-precision distance calculation is executed on a specialized accelerator featuring bit-serial compute units and a bit-interleaved data layout, enabling efficient variable-precision execution. In addition, we design a greedy scheduling strategy to balance workloads across compute units operating at different precisions.

In summary, we make the following contributions:
\begin{itemize}
\item We propose ANNS-AMP, an adaptive mixed-precision ANNS framework and accelerator that supports general cluster-based methods, including PQ and its variants, across flexible bit-width configurations.

\item We develop a precision prediction scheme based on cluster-level feature extraction, enabling adaptive precision scaling with minimal accuracy loss.

\item We design a hardware architecture with bit-serial computation and bit-interleaved data layout, and address key challenges such as bandwidth bottlenecks and load imbalance through tailored scheduling strategies, with efficient reuse of the bit-serial computing array for runtime precision prediction.

\item We implement ANNS-AMP using a cycle-accurate simulator and evaluate it on representative datasets. Results show that ANNS-AMP achieves up to 163.76$\times$, 10.57$\times$, and 2.06$\times$ performance speedups on average, while reducing average energy consumption by 1100.00$\times$, 39.41$\times$, and 6.66$\times$ compared to CPU, GPU, and ASIC baselines, respectively. Notably, up to 87.49\% and 93.75\% of distance computations in the cluster location and LUT construction phases are performed in low precision, with overall accuracy loss below 2.7\%.
\end{itemize}

\section{Background and Motivation}

\subsection{Characterization of Typical ANNS}

Cluster-based indices are widely adopted for ANNS due to their sequential memory access patterns and compact storage \cite{refp10, refp06, refp05, refp11, refp04, refp09, refp12, refp13}. In such indices, the vector corpus is partitioned into clusters, and each query probes only its nearest clusters. As clustering reduces the effective search space, vectors are further compressed using product quantization (PQ) with a shared codebook across clusters. Specifically, PQ partitions the original vector space into $M$ subspaces along the dimensions, clusters the residuals between vectors and their corresponding cluster centroids within each subspace to form codebooks, and encodes each vector as a combination of codebook entry indices across subspaces.

\begin{figure*}[!t]
\centering
\includegraphics[width=6.5in]{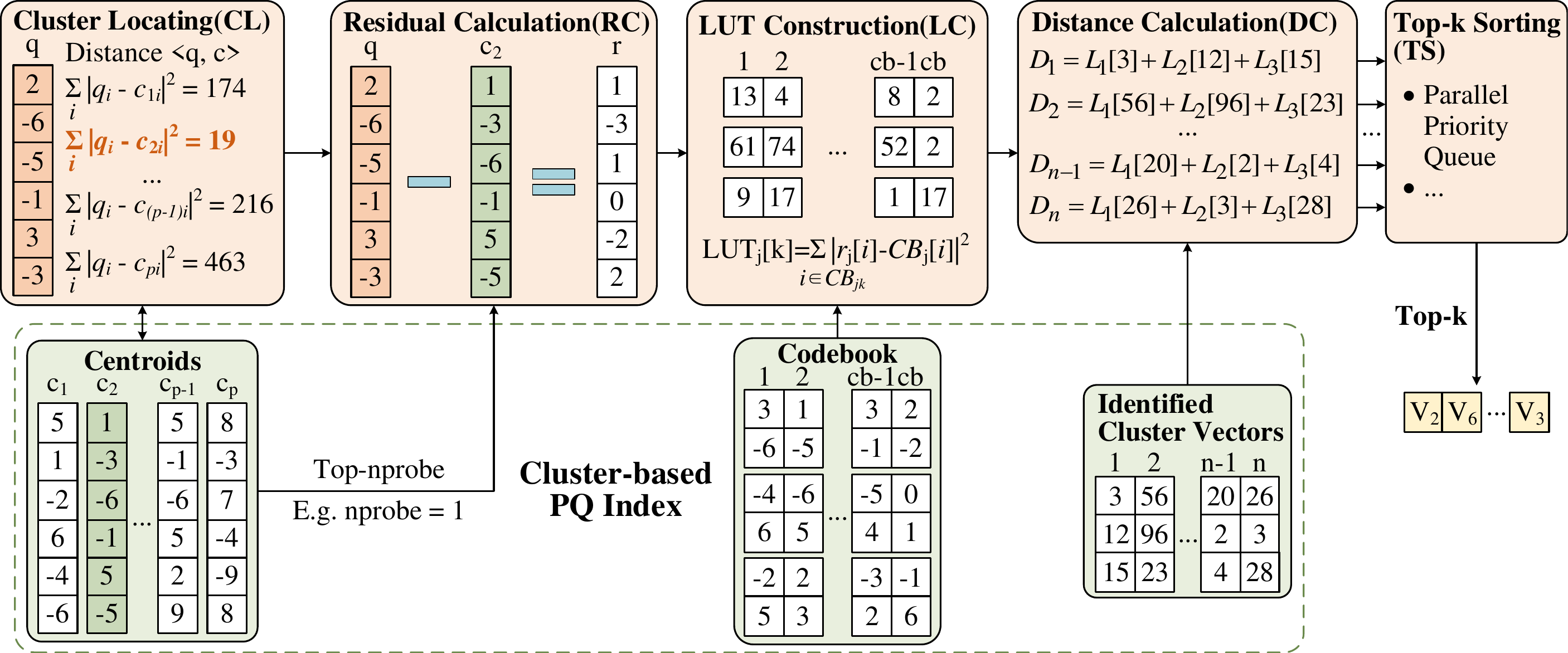}
\caption{Illustration of ANNS with cluster-based PQ index (L2 distance).}
\label{fig1}
\end{figure*}

As illustrated in Figure~\ref{fig1}, ANNS with a cluster-based index typically consists of five stages: cluster locating (CL), residual calculation (RC), LUT construction (LC), distance calculation (DC), and top-$k$ selection (TS). In CL, each query exhaustively searches for its nearest cluster centroids based on the L2 distance. For the selected clusters, RC computes the residual between the query and each cluster centroid. Then, in LC, distances between the residual and codebook entries are computed in each of the $M$ subspaces to construct lookup tables (LUTs). In DC, these LUTs are indexed by encoded vectors and accumulated to estimate the distances between the query and database vectors within the clusters. Finally, TS ranks the computed distances to retrieve the top-$k$ nearest neighbors.

Among these stages, distance computations in CL and LC account for a substantial portion of the overall overhead \cite{refp06, refp09}. Notably, the majority of these computations involve non-neighbor vectors, constituting approximately 99.94\% and 99.88\% of operations in CL and LC, respectively, under typical ANNS configurations. To mitigate such redundancy, prior works have explored early termination at the vector level (ETV) \cite{refp01, refp03}, dimension level (ETD) \cite{refp45}, and bit level (ETB) \cite{refp43, refp24}. However, these approaches rely on maintaining a top-\emph{k} list for pruning and thus become less effective in stages such as LC, where distance computations are used to construct lookup tables rather than directly support top-\emph{k} selection. As shown in Table~\ref{table1}, under high-accuracy configurations of cluster-based ANNS, more than 51.86\% of redundant computations remain after applying these techniques. Moreover, vector-level early termination strategies are often tailored to graph- or tree-based indices, leading to significant accuracy degradation when applied to cluster-based structures.

To reduce redundancy in LC, JUNO \cite{refp05} partitions the codebook into grids and prunes distance computations between the query and codebook entries located in distant grids. The pruning threshold is determined via a polynomial regression model to preserve over 90\% of the top-100 candidates. However, due to the limitations of grid-based partitioning in high-dimensional spaces and the resulting incompleteness of the constructed LUTs, a substantial gap remains between this approach and the theoretical upper bound of redundancy reduction. Other approaches \cite{refp06, refp02, refp09} rely on static parameter tuning, which lacks adaptability to variations in query distributions.

To further exploit the substantial redundancy in distance computations, we investigate adaptive mixed-precision computing to assign appropriate precision to each distance calculation. Specifically, we employ a regression model to predict precision requirements at the cluster level and adopt bit-serial computation to enable efficient mixed-precision execution at runtime.

\subsection{Bit-serial Computation}

Bit-serial computation processes one bit per cycle. For instance, a bit-serial multiplier requires only an AND gate and a full adder per bit position: the corresponding bits of the two operands are fed to the AND gate, and the result is input to the adder along with the carry from the previous bit and the sum from the less significant bit position. Consequently, under the same area and power constraints, bit-serial computational units achieve throughput comparable to that of multi-bit processors \cite{refp15}, and their throughput scales inversely with the operand bit-width. As shown in Table~\ref{table1}, the redundancy gap between the theoretical upper bound and early termination based on a top-\emph{k} list with bit-serial computing (ETB) remains substantial, indicating further potential for precision prediction with higher accuracy and avoiding dependence on the top-\emph{k} list for thorough redundancy reduction. Moreover, current ETB fails to improve bandwidth utilization, since its pruning is irregular across elements and operates at a granularity far finer than that of memory accesses. 
Therefore, despite the dynamic reconfigurable precision provided by bit-serial modules, ANNS acceleration via adaptive mixed-precision arithmetic is still confronted with several challenges. First, while ETB at element level is limited by its dependence on the top-\emph{k} list and yields negligible bandwidth improvement, conventional grid-based partition scales poorly in high-dimensional vector spaces due to the ineffectiveness of hyperplane-based division in such spaces \cite{refp14, refp11}. Meanwhile, non-neighbor vectors are not consistently far from each other in every 2-dimensional grid-based slice, limiting the accuracy of precision prediction with this partition scheme. Second, in the absence of the top-\emph{k} list in some phases, the prediction model requires other representative features to adapt distance calculation between queries and ANNS index data to appropriate precision levels. Third, ANNS is both compute-intensive and memory-intensive, and may become bounded by bandwidth depending on query and base vector distributions, necessitating a specialized vector layout that coalesces memory requests for data sharing the same precision to improve bandwidth efficiency. In addition, the latency of bit-serial units varies with the computation precision, thus requiring load scheduling to balance the dynamic tasks of distance calculation.

\begin{table}[t!]
  \centering
  \caption{Redundancy reduction of distance calculation in prior ANNS works.}
  \label{table1}
  \begin{threeparttable}
      \begin{tabular}{cccccc}
        \hline
         & ETV \cite{refp01} & ETD \cite{refp45} & ETB \cite{refp24} & JUNO \cite{refp05} & Upper-bound\\
        \hline
        CL & 99.60 \% \tnote{*} & 52.15 \% & 72.08 \% & 0 & 99.94\%  \\
        LC & 0 & 0 & 0 & 64.84 \% & 99.88 \% \\
        All & 66.40 \% & 34.77 \% & 48.05 \% & 21.61 \% & 99.92 \% \\
        \hline
      \end{tabular}
      \begin{tablenotes}
        \footnotesize
        \item[*] The accuracy loss exceeds 80\%.
      \end{tablenotes}
  \end{threeparttable}
\end{table}

\section{ANNS-AMP Framework}

\subsection{Overview}

\begin{figure}[!t]
\centering
\includegraphics[width=3.3in]{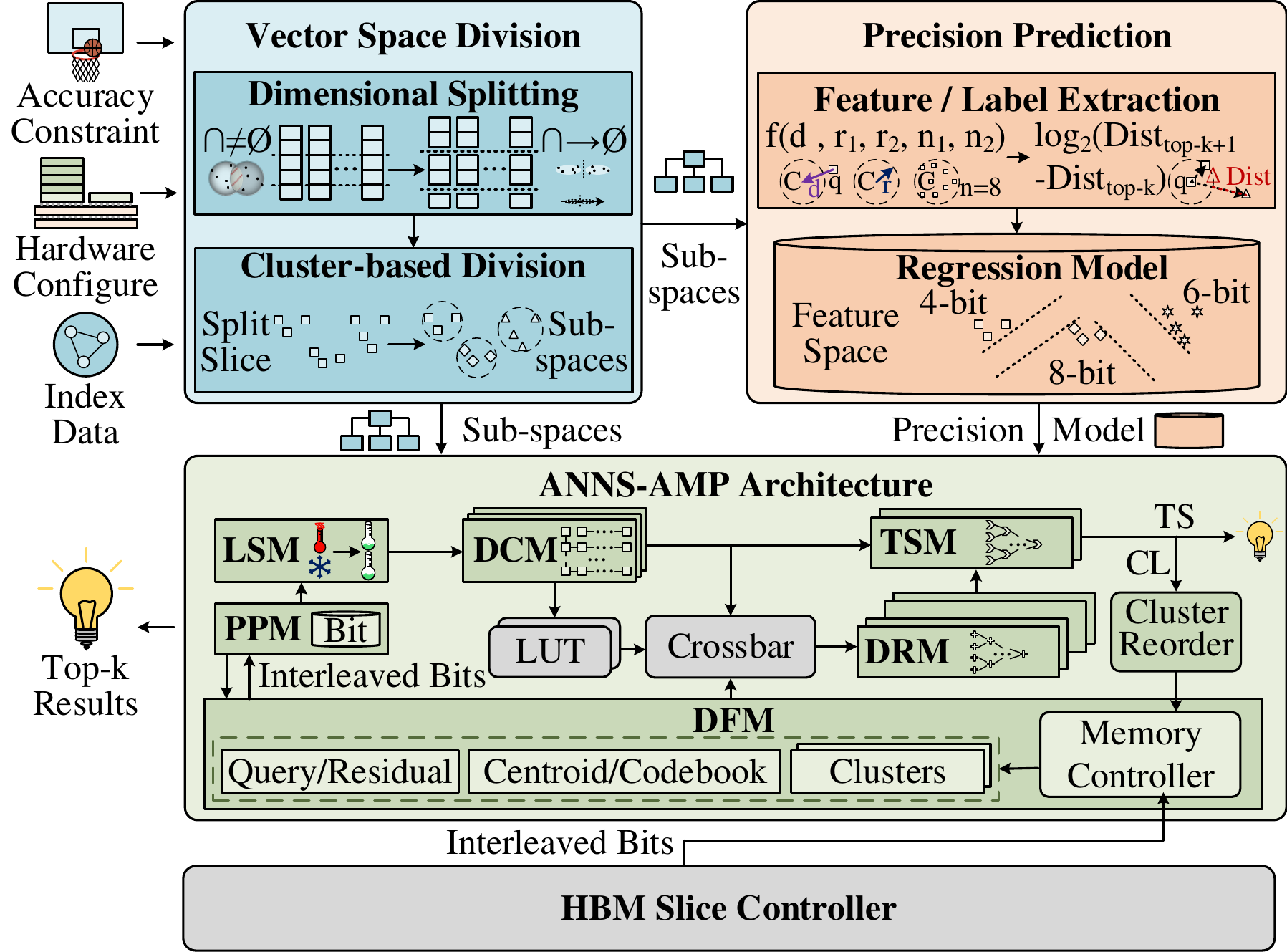}
\caption{The overview of ANNS-AMP framework.}
\label{fig3}
\end{figure}

In this work, we propose ANNS-AMP, an ANNS accelerator via adaptive mixed-precision computing. The framework co-designs precision prediction algorithm and hardware precision fitting for ANNS. As shown in Figure~\ref{fig3}, ANNS-AMP firstly introduce space division in both dimension level and vector level to obtain independent sub-spaces. In each sub-space, it extracts representative features which reflects the tolerant hardware precision. The features are utilized to train a support vector regression (SVR) model offline. At runtime, the precision of each sub-space is predicted by the SVR model based on the local features and the location of the query. To convert the benefits from the mixed-precision algorithm optimizations to efficient hardware implementation, ANNS-AMP proposes a precision-adjustable accelerator with optimizations in memory and scheduling based on specific workload analysis.

\subsection{Vector Space Division}
\label{sec:vector_space_division}

\begin{figure}[!t]
\centering
\includegraphics[width=2in]{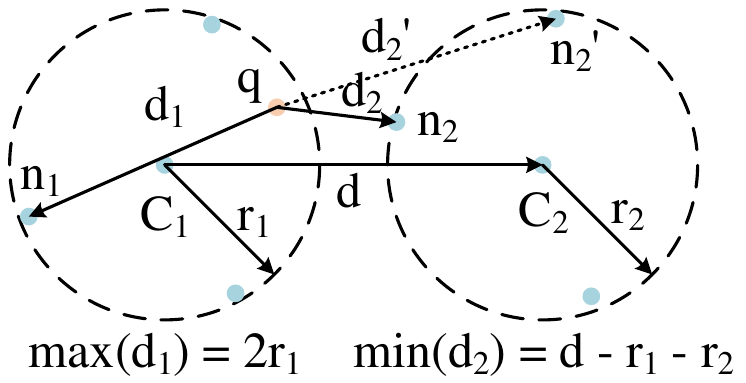}
\caption{Illustration of the relation among a query, the sub-space it locates, and the neighbor sub-space.}
\label{fig4}
\end{figure}

In a $D$-dimensional space, sub-spaces divided by hyperplanes have $O(D)$ neighbors for each. For example, a 2-dimensional grid has 4 neighbor grids, and a 3-dimensional cube has 6 neighbor cubes. Since a query might locate near the border of a sub-space, the neighbor sub-spaces or neighbors of neighbors are candidates for search as well. Therefore, high dimension increases the burden of neighbor searching. Cluster-based division divide vector space by hyperspheres, allowing intersection among sub-spaces. However, frequent intersection would also cause costly search in multiple neighbor sub-space, which is exacerbated in high-dimensional vector space. As illustrated in Figure~\ref{fig4}, assume the distance between cluster $C_1$ and $C_2$ is $d$, the radii of $C_1$ and $C_2$ are $r_1$ and $r_2$ respectively, and the distance between the query $q$ and potential neighbors of it in $C_1$ and $C_2$ is $d_1$ and $d_2$, then the probability that the potential neighbor $n_1$ in $C_1$ is more closed to $q$ than the potential neighbor $n_2$ in $C_2$ is:

\begin{equation}
\label{eq1}
    P(d_1 \le d_2) \ge P(2r_1 < d - r_1 - r_2)
    = P(d > 3r_1 + r_2)
\end{equation}

\begin{figure}[!t]
\centering
\includegraphics[width=3in]{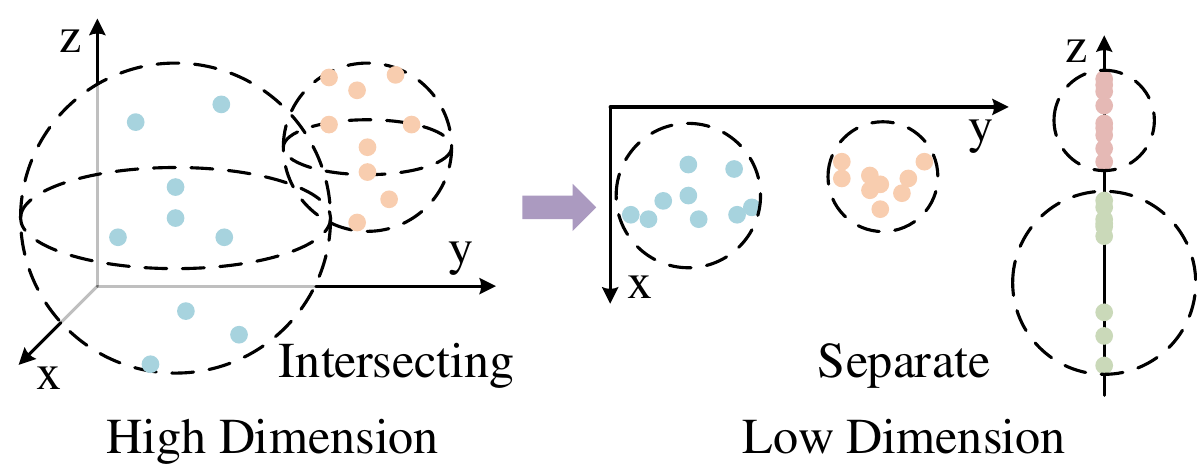}
\caption{Intersecting high-dimensional sub-spaces may be separate in low-dimensional projection space.}
\label{fig5}
\end{figure}

\begin{figure}[!t]
\centering
\includegraphics[width=3.3in]{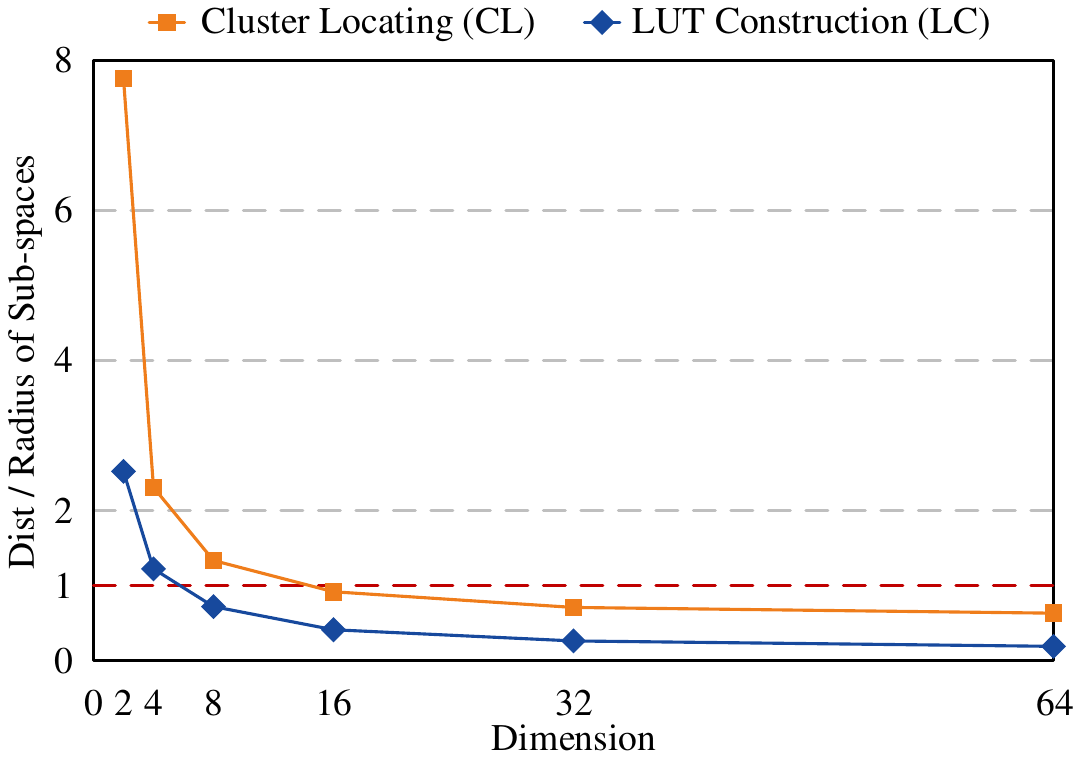}
\caption{Average ratio between neighbor sub-space distances and the sum of their radius in CL and LC on SIFT100M. $N$ input vectors are split into $\sqrt{N}$ sub-spaces. The vertical axis represents the density of sub-spaces.}
\label{fig2}
\end{figure}

\noindent
Assume that $r_1 = r_2 = r$, then the lower bound of the probability above is $P(\frac{d}{r} > 4)$. Therefore, it would reduce the overhead of search neighbor sub-spaces to improve the sparsity of sub-spaces or shrink the size of each sub-space. Since the size of vector corpus is fixed, the number, or the sparsity, of sub-spaces is also stable. However, as illustrated in Figure~\ref{fig5}, intersecting high-dimensional sub-spaces may become separate in low-dimensional projection space due to reduction of $r$, which is proved practically as shown in Figure~\ref{fig2}.

On the other hand, neighbor vectors in high-dimensional space might not be closed in some dimensions. When applied mixed-precision acceleration, the loss of precision in these dimensions would hurt the accuracy of distance calculation for neighbors, which generally necessitates accurate guarantees to be distinguished by other vectors. Meanwhile, fine-grained vector division brings loading and storage overheads of cluster mapping, as will be introduced in Section~\ref{sec:hardware:memory}. Therefore, in the offline space division phase, we empirically select the dimension of sub-spaces to exactly make most sub-spaces separate from their neighbor sub-spaces based on Equation~\ref{eq1} while keeping accuracy information in high-dimensional space.

\subsection{Sub-space Feature Extraction}
\label{sec:feature_extraction}

In the divided sub-spaces, the calculated distance would be accumulated to form the complete one. Therefore, each partial distance between the query and sub-spaces cannot be pruned even if they are far. To quantify the relation between the partial distance and the corresponding precision, suppose the condition in Equation~\ref{eq1} as:

\begin{equation}
\label{eq2}
    f(d, r_1, r_2) = d - 3r_1 - r_2
\end{equation}

\noindent
As illustrated in Figure~\ref{fig4}, the precision required by $C_2$ reduces with the increase of $f(d, r_1, r_2)$. Meanwhile, since the amount of top-\emph{k} neighbors is given, the amount of vectors contained in each sub-space is also a factor of required precision. Therefore, the precision required by $C_2$ can be expressed by $f(d, r_1, r_2, n_1, n_2)$, where $C_1$ is the closest sub-space of $q$, and $n_1$ and $n_2$ represent the amount of vectors in $C_1$ and $C_2$ respectively.

However, calculation and storage of $d$ are costly since the complexity is $O(D \times C^2)$, where $C$ is the amount of sub-spaces in a divided $D$-dimensional space. On the other hand, it requires the distance $d'$ between $q$ and each sub-space to locate $C_1$ that is nearest to $q$ in each divided $D$-dimensional space. Therefore, we replace $d$ by $d'$ to form a combination of features $\{d', r_1, r_2, n_1, n_2\}$ for precision prediction.

\begin{figure}[!t]
\centering
\includegraphics[width=3in]{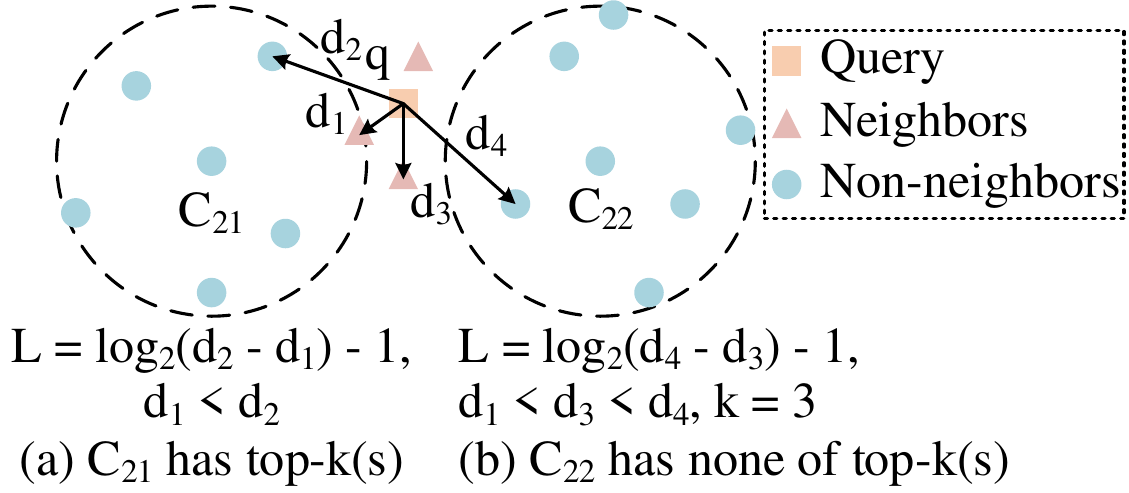}
\caption{Illustration of label generation. (a) The sub-space contains global neighbor(s). (b) The sub-space contains none of global neighbors.}
\label{fig6}
\end{figure}

Since the format of the precision function $f(d', r_1, r_2, n_1, n_2)$ is implicit, we train a regression model to fit it, which will be introduced in Section~\ref{sec:precision_prediction}. Labels for the supervised learning is obtained by the difference between the ground-truth distance from $q$ to global neighbors and non-neighbors in each sub-space, or the difference between the ground-truth distance from $q$ to the local nearest vector in a sub-space where no global neighbors locate, and the furthest global neighbor. An example is illustrated in Figure~\ref{fig6}. The target of labels is to distinguish global neighbors in each space from those non-neighbors with a precision as low as possible. The features and labels are generated on a profiled ground-truth set.

\subsection{Precision Prediction}
\label{sec:precision_prediction}

We adopt a classic regression model, support vector regression (SVR) \cite{refp16}, to fit the precision function $f(d', r_1, r_2, n_1, n_2)$. The model distributes samples to each kind of precision by borders with maximal distance. Though the $f(d, r_1, r_2)$ presented by Equation~\ref{eq2} is linear, the relation between the precision and $n_1$ and $n_2$ is implicit. Therefore, we utilize a radial basis function (RBF) as the kernel function of SVR. Specifically, we utilize Gaussian kernel for the feature mapping to a linear space. During the online inference phase, the results of the non-linear function is obtained by a look-up table to avoid exponential and division operations. As for training samples, we picked up only 1280 pairs of features and labels due to the computational and memory complexity of SVR which is the square of sample amount. The model is utilized to predict the precision of sub-spaces at runtime based on the location of each query.

\section{ANNS-AMP Micro-architecture}

\begin{figure}[!t]
\centering
\includegraphics[width=3.3in]{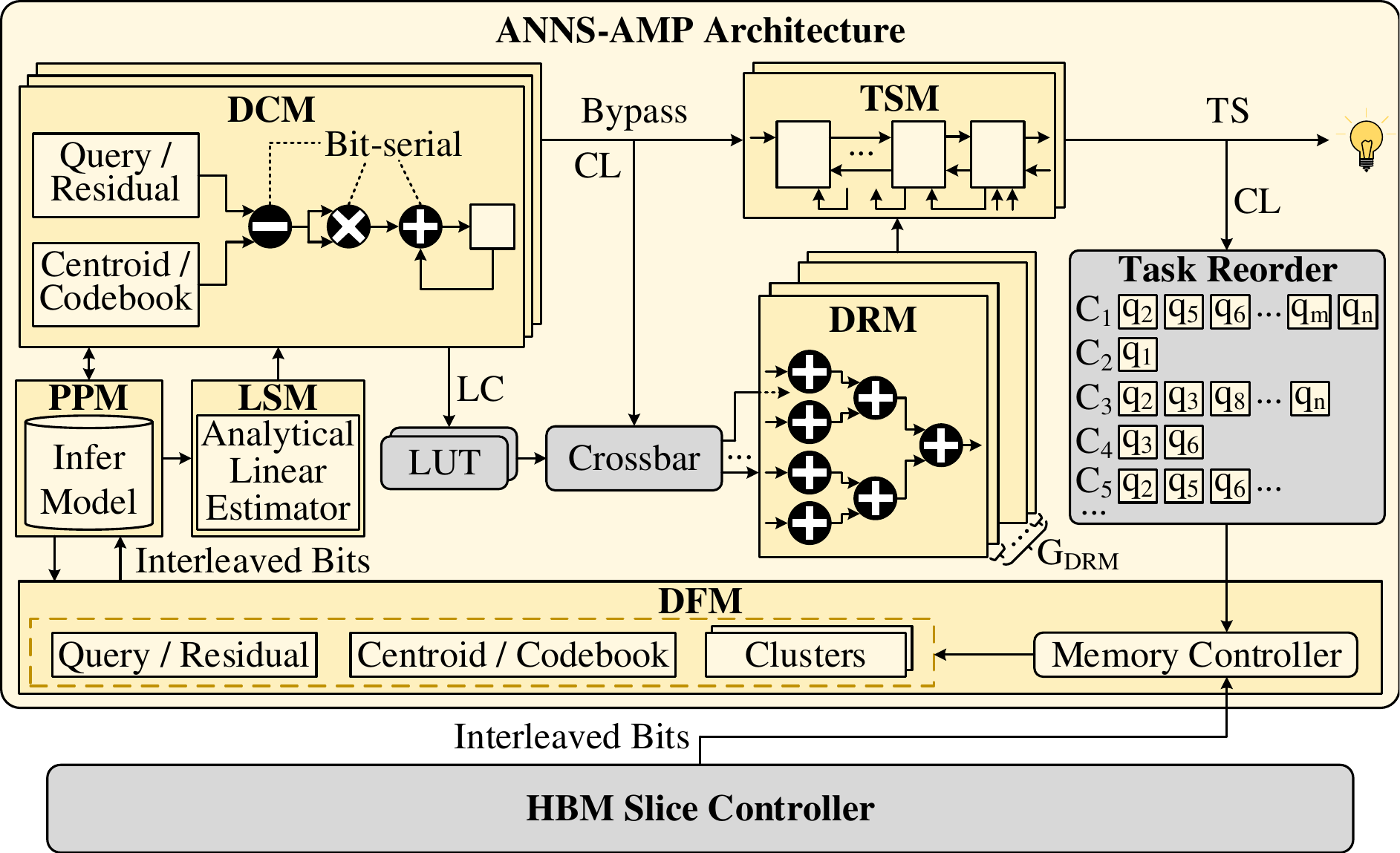}
\caption{The overall hardware architecture of ANNS-AMP.}
\label{fig7}
\end{figure}

Vector space division and precision prediction make it possible to accelerate distance calculation in sub-spaces requiring low precision. However, general processors have limited supports for low-precision computation. To maximize the benefits of mixed-precision acceleration, we design a specific accelerator, ANNS-AMP, based on bit-serial optimization. Since some ANNS phases is bound by memory bandwidth instead of computation when the given accuracy constraint changes, we apply mixed-precision layout optimizations to the ANNS-AMP for further acceleration. Unlike multi-bit calculation with the same precision, low-precision operations is faster than high-precision ones, leading to load imbalance across the computing array. Therefore, we propose a greedy scheduling strategy based on an analytical model for performance estimation. The overall hardware architecture is shown in Figure~\ref{fig7}.

\subsection{Micro-architecture Overview}

The computing engine of ANNS-AMP contains four modules: Distance Calculation Module (DCM), Distance Reduction Module (DRM), Top-k Sorting Module (TSM), and Precision Prediction Module (PPM). To optimize memory accesses and load balance on ANNS-AMP, we further design a Data Fetching Module (DFM) and Load Scheduling Module (LSM), which will be introduced in Section~\ref{sec:hardware:memory} and Section~\ref{sec:hardware:sched} respectively.

\textit{Distance Calculation Module (DCM):} As shown in Figure~\ref{fig7}, DCM calculates L1/L2 distance of input vectors in a single dimension, including a subtractor, a multiplier, and an adder. All of these computational units are bit-serial. DCM can support distance calculation in CL, RC, and LC by reusing components of it. 
In CL, all of the three components in DCM are utilized, where slices of the query and centroid are input to the subtractor to obtain L1 distance at first, and the multiplier calculates the square to obtain L2 distance in one dimension. The 1-dimensional distance is accumulated by the adder to obtain the partial distance between the input slices. The operations in the independent components are implemented in pipeline. The process in LC is similar, where the inputs are replaced by slices of the residual and codebook. As for RC, it only utilizes the subtractor and bypasses other components to obtain the residual between the query and centroid.

\textit{Distance Reduction Module (DRM):} As shown in Figure~\ref{fig7}, DRM accumulates the partial results of distance in CL, LC and DC by an adder tree. In CL and LC, DRM will be bypassed if the vector space is not divided. In DC, the inputs come from a distance LUT constructed in LC. The adders are multi-bit in DRM to simplify the implementation. Therefore, the inputs are extended to the same bit width before accumulation. To meet the demands of high parallelism of bit-serial components in DCM, we set multiple DRMs for ANNS-AMP, which are divided into group and each group processes the workloads of one query, connecting with a distance LUT and a top-\emph{k} priority queue.

\textit{Top-k Sorting Module (TSM):} TSM contains a priority queue for top-\emph{k} sorting. As shown in Figure~\ref{fig7}, TSM is connected with DRM to sort distance results in CL and DC. The top results is forwarded to the result buffer of DRM for pruning. The initial values of the priority queue are set as maximum to prevent pruning before the priority queue is full.

\textit{Precision Prediction Module (PPM):} PPM utilizes a pretrained regression model to predict the precision of each sub-space. The size and radius of each sub-space are calculated offline and loaded to form input features. DCM and TSM are reused to locate the nearest sub-space to the query and obtain the distance between the query to each sub-space, which help to form the features as well, as mentioned in Section~\ref{sec:feature_extraction}. The precision of each sub-space obtained by PPM is transferred to DCM, DFM, and LSM for acceleration.

\subsection{Memory Optimizations}
\label{sec:hardware:memory}

\begin{figure}[!t]
\centering
\includegraphics[width=3in]{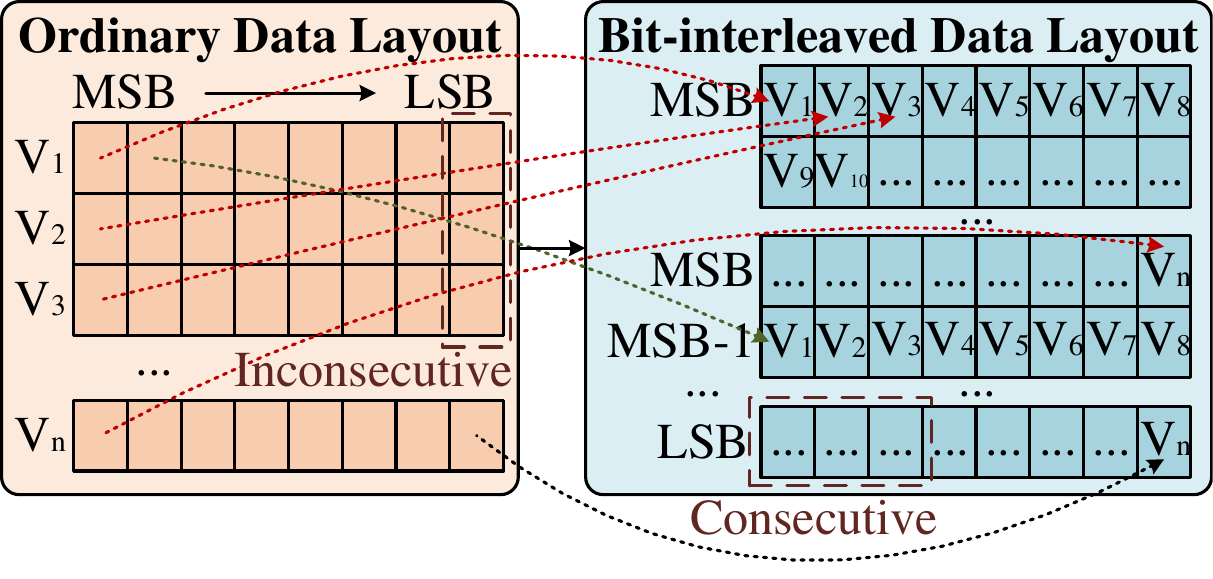}
\caption{Illustration of the bit-interleaved memory layout.}
\label{fig17}
\end{figure}

The compute-to-IO ratio is changed by mixed-precision distance calculation due to the reduction of computation. Therefore, the bottleneck might change to memory accesses without memory optimizations. To benefit data loading from mixed-precision computation, we adopt a bit-interleaved memory layout for queries, centroids and codebook that will be input to mixed-precision modules. Vectors in each sub-space are stored in serial. As illustrated in Figure~\ref{fig17}, based on the bit-interleaved memory layout, the most significant bit (MSB) of each kind of these data are stored serially, and the same is for the bit after MSB, until the least significant bit (LSB). With the bit-interleaved layout, the computational modules only access the corresponding bits of a sub-space predicted by PPM in serial. ANNS-AMP equips a Data Fetching Module (DFM) to fetch data, which loads the bits to a ping-pong buffer and sends them to the corresponding computational module in order. DFM supports prefetching to overlap data loading with computing by the ping-pong buffer.

\subsection{Scheduling Optimizations}
\label{sec:hardware:sched}

The latency of each computational units depends on its precision. Meanwhile, as mentioned in Section~\ref{sec:vector_space_division}, the size of each sub-space depends on the distribution of vectors, aggravating load imbalance across sub-spaces, since DCMs cannot process exactly one sub-space at all time, leading to idle units when processing by sub-spaces. To alleviate the load imbalance, we introduce paths connecting neighbor DCM groups. When a DCM group becomes idle, it would send a signal to its busy neighbor group to offload computation greedily. Since we only connect DCM groups when they are neighbors, the overheads of these paths are negligible. We apply a simple analytical predictor to estimate the workloads of each sub-space based on its size, dimension and precision for task allocation, which further alleviates performance loss caused by task offloading across neighbor groups.

\section{Evaluation}

\subsection{Experiment Setup}
\label{sec:evaluation:setup}

We implemented an in-house cycle-accurate simulator for performance measurement of ANNS-AMP and an RTL design synthesized with Synopsys design compiler (DC) in TSMC 14-nm process technology. The accelerator is built on the logic layer with a 3-D stacked memory module which is divided into 32 pseudo channels with a total internal bandwidth of 1600 GB/s. The frequency of the accelerator is set as 1 GHz. Similar to the method of power estimation utilized in \cite{refp17}, we estimated the power of logic modules using Synopsys PrimeTime by annotating the switching activity. The performance and the power of the 3-D memory are estimated by ramulator \cite{refp18} and the simulator proposed in \cite{refp19}, respectively.

We compared ANNS-AMP with state-of-the-art open-sourced ANNS implementations including Faiss-CPU, Faiss-GPU \cite{refp04} and a typical ANNS accelerator design, ANNA \cite{refp10}. CPU platform equips 32-thread Intel Xeon Gold 5218@2.3GHz processor and 512GB DDR4. GPU platform equips an NVIDIA A100 PCIe and 80GB HBM2e whose bandwidth is 1935GB/s. The processors in CPU and GPU platform utilize 14-nm and 7-nm process technology respectively, which are similar to or better than that of ANNS-AMP. Faiss-CPU is accelerated by both AVX instructions and multi-thread optimizations. ANNA \cite{refp10} is based on cluster-based index and works at the same frequency as ANNS-AMP. For a fair comparison, we utilized the 12-instance version, ANNAx12, and limited the bandwidth of ANNS-AMP to 800GB/s during the comparison with ANNAx12 so that both of them have the same bandwidth. Since ANNA is not open-sourced, we built a simulator based on its designs with the same process technology as ANNS-AMP and verified the performance by typical results in its paper. The energy of CPU and GPU is obtained from Intel RAPL domain \cite{refp20} and nsys respectively. To avoid data overflow during comparison with GPU, we utilized the 100-million-vector version of both SIFT \cite{refp21} and DEEP \cite{refp22} datasets, i.e. SIFT100M and DEEP100M. The dimension of them are 128 and 96 respectively. DEEP100M is quantified to uint8 to keep in coincidence with SIFT100M. The query set of both datasets includes 10000 queries. The amount of samples for training the regression model is limited to a maximum of 1280, and the iterations are no more than 50. Due to the different features between the cluster locating phase (CL) and LUT construction phase (LC), we alternative specific hyperparameters in each phase. In CL, $\gamma$ and $C$ are set as 0.1 and 10 respectively. In LC, both $\gamma$ and $C$ are changed to 1 to avoid overfitting due to lack of sub-spaces. According to prior works \cite{refp10, refp06, refp23, refp24, refp09}, the default accuracy constraint is set as recall@10 $\ge$ 0.8.

\subsection{Performance Analysis}
\label{sec:evaluation:overall}

\begin{figure}[!t]
\centering
\includegraphics[width=2.5in]{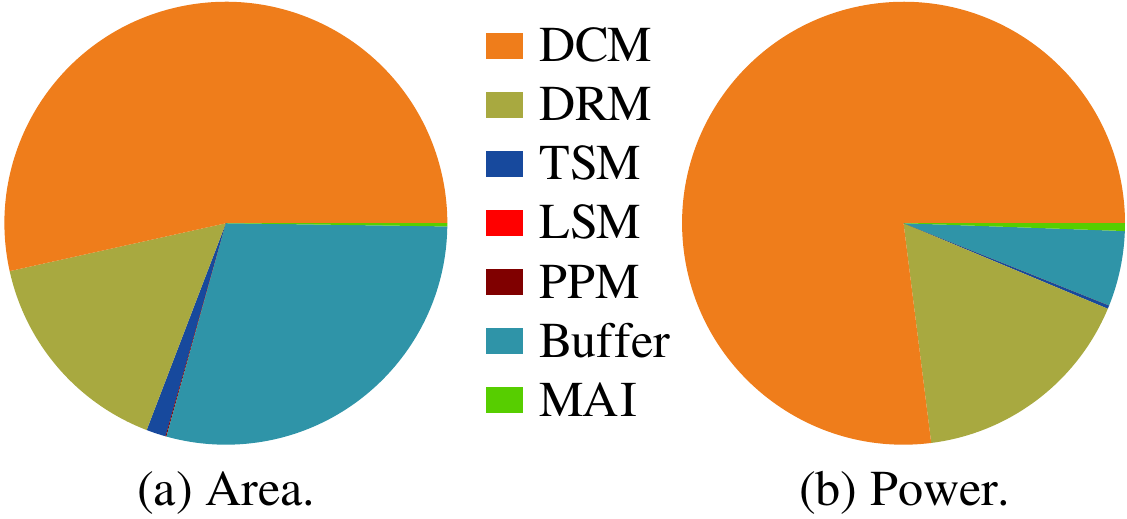}
\caption{The area and power breakdown graphics.}
\label{fig9}
\end{figure}

\textbf{Power and area:} ANNS-AMP sets 1024 groups of bit-serial processing units in DCM. Each group contains 32 processing units for parallel distance calculation of 32-dim vector slices. Thus, $N_{DCM}=1024 \times 32=32768$. DRM is configured with 1024 processing units, i.e. $N_{DRM}=1024$. The units are divided to 4 groups to process hot and cold clusters in an interleaved way with prefetching, i.e. $G_{DRM}=4$. Each unit in DRM contains a 32-input adder tree and processes each level by pipeline. TSM is triggered by DCM or DRM, and it equips 1024 priority queues to meet the demands of task parallelism in DCM and DRM. PPM reuses DCM for distance calculation to obtain features in sub-spaces and predicts the precision based on the features with the regression model by itself. Similarly, LSM utilizes global load features for scheduling, so it is pipelined with PPM. Cluster buffer is set as 1MB and supports prefetching 32768 16-dim PQ-encoded vectors at most. If clusters exceed the buffer, they would be loaded by stream. Codebook buffer is set as 64KB and can accommodate 256 256-dim codebook at most. The size of buffers for centroids, queries and residuals is set as 256KB, which is available to 2048 128-dim vectors or 1024 256-dim vectors at most. In summary, the total amount of on-chip buffers is 1.8125MB. All of them are implemented by SRAMs. Figure~\ref{fig9} presents the area and power breakdown, where MAI represents memory access interface. Total chip area is 7.549mm$^2$ and the power consumption is 11.451W in TSMC 14-nm technology.

\begin{figure}[!t]
\centering
\includegraphics[width=3.3in]{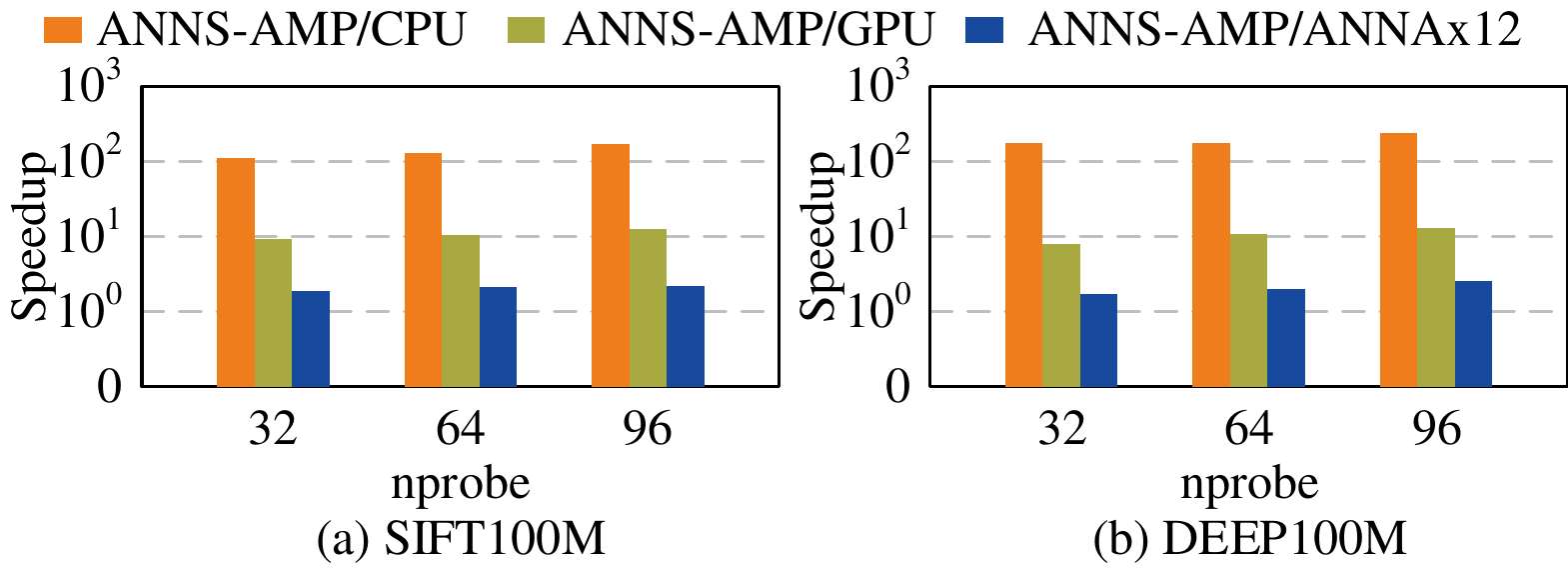}
\caption{Performance speedup over CPU, GPU, ANNAx12.}
\label{fig10}
\end{figure}

\textbf{Performance:} We compare the performance of ANNS-AMP with Faiss-CPU, Faiss-GPU and ANNAx12 with the same algorithm configurations. DEEP100M is quantified to uint8 for Faiss as well. Performance is measured by throughput in queries per second (QPS). For a fair comparison, we restrict the bandwidth of ANNS-AMP to 800GB/s when compared with ANNAx12 so that both of them have the same bandwidth. As shown in Figure~\ref{fig10}, ANNS-AMP is 163.76 $\times$, 10.57 $\times$, and 2.06 $\times$ faster than Faiss-CPU, Faiss-GPU, and ANNAx12 respectively. Specifically, ANNS-AMP shows higher speedup on DEEP100M, whose dimension is not a power of two, over CPU. The reason is that the granularity of vector processing on CPU is set as a multiple of 64 by the baseline, causing a waste of computational power in CL on DEEP100M. Compared to them, ANNS-AMP processes vectors at a fine-grained granularity and is less sensitive to dimension. In addition, ANNS-AMP exhibits higher speedup over CPU and GPU with larger $nprobe$, where each query retrieves more clusters, indicating better adaptability of strict accuracy constraints. Compared to ANNAx12, ANNS-AMP benefits from mixed-precision distance calculation and corresponding layout optimizations, obtaining better performance even with the same bandwidth configuration.

\begin{figure}[!t]
\centering
\includegraphics[width=3.3in]{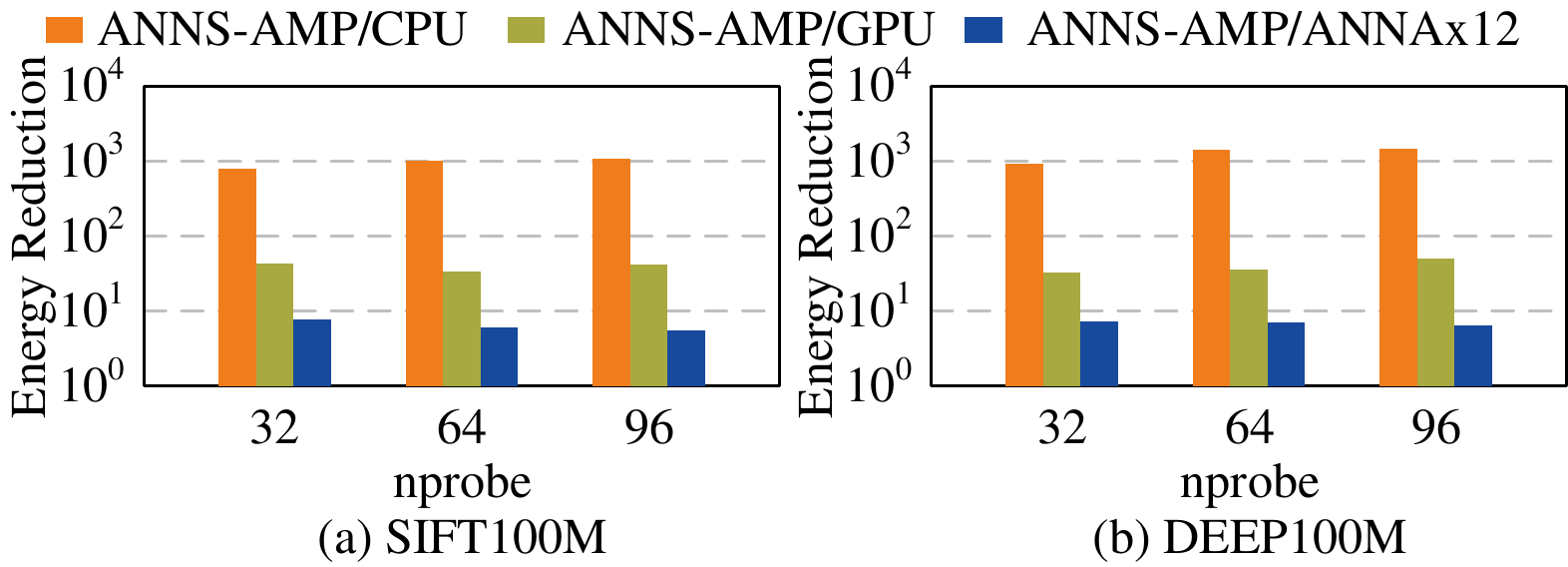}
\caption{Energy reduction over CPU, GPU, ANNAx12.}
\label{fig11}
\end{figure}

\textbf{Energy:} As shown in Figure~\ref{fig11}, ANNS-AMP achieves about 1100.00 $\times$, 39.41 $\times$, and 6.66 $\times$ less energy consumption compared with CPU, GPU, and ANNAx12 respectively. Compared to CPU and GPU, on top of the execution advantage, the energy consumption reduction is mainly attributed to the unified customized accelerator design with less redundancy in computational modules and specific data reuse. In addition, ANNS-AMP exhibits energy reduction over ANNAx12 even with the same bandwidth thanks to mixed-precision acceleration that saves power and area while keeping the performance. Meanwhile, ANNS-AMP adopts the bit-interleaved memory layout to improve the bandwidth efficiency of HBM without naively adding instances, contributing to the energy reduction while improving throughput.

\subsection{Optimizations}

\begin{figure}[!t]
\centering
\includegraphics[width=3.3in]{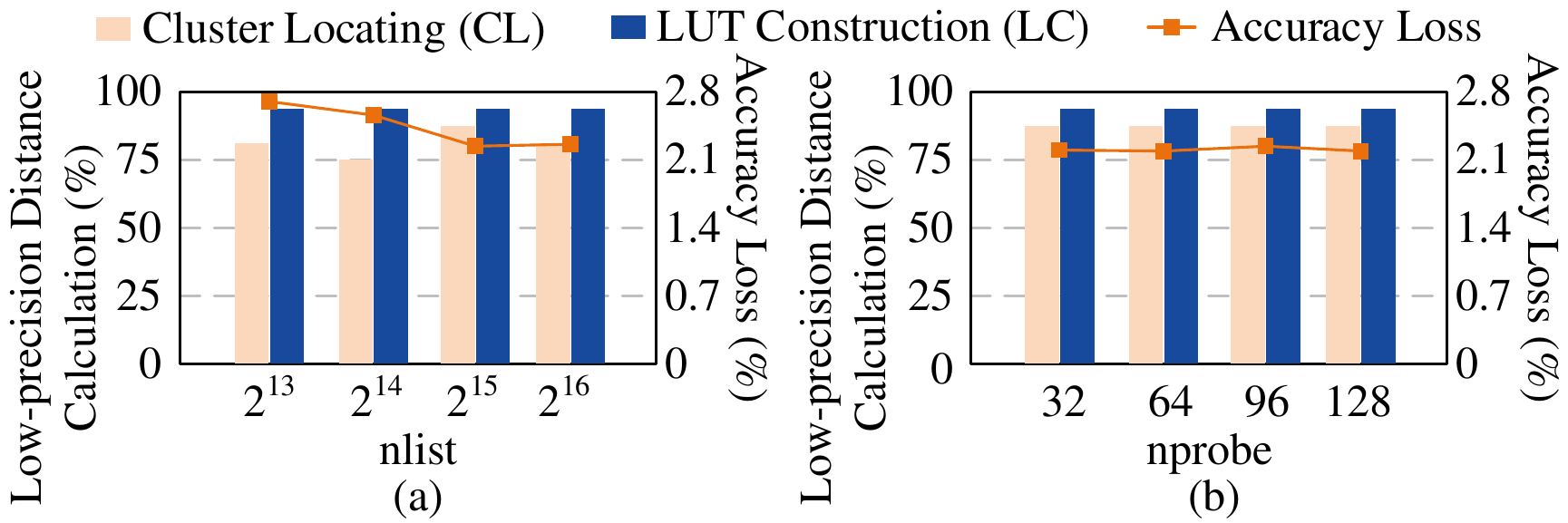}
\caption{Accuracy loss with mixed precision ANNS among various parameter configurations on SIFT100M.}
\label{fig12}
\end{figure}

\textbf{Precision prediction:} Figure~\ref{fig12} shows the effects of mixed-precision computation on accuracy across various index parameters on SIFT100M. In Figure~\ref{fig12}(a), the amount of clusters $nlist$ varies with other parameters fixed, while only $nprobe$ changes in Figure~\ref{fig12}(b). The bars present the ratio of low-precision distance calculation in CL and LC, and the plot presents overall accuracy loss caused by mixed-precision computation. Across all typical index parameters, 74.98\%-87.49\% of distance calculation benefits from low-precision processing in CL, while the ratio increases to more than 93.75\% in LC, though the original index vectors in CL and LC have already been quantified to a 8-bit representation. Meanwhile, the overall accuracy loss is kept below 2.7\%. Specifically, in CL, the errors caused by precision prediction would make ANNS miss near vectors at a granularity of clusters, leading to a stricter demands of precision prediction than other phases. Therefore, CL benefits less from low-precision processing than LC. Nonetheless, in some cases, the difference of utilization in low-precision processing between CL and LC shrinks to less than 6.251\%. In summary, distance calculation across different ANNS phases can benefit from mixed-precision acceleration with a bit accuracy loss using typical index parameters, indicating the effectiveness of precision prediction.

\begin{figure}[!t]
\centering
\includegraphics[width=3.3in]{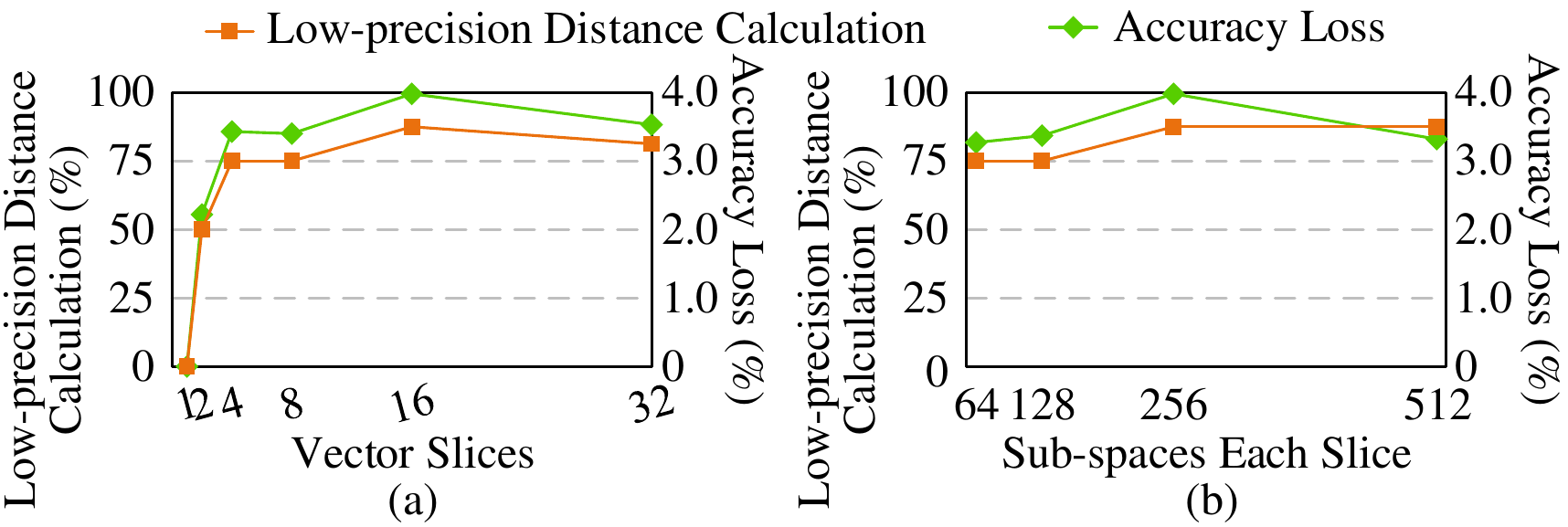}
\caption{Benefits from mixed precision ANNS among various sub-space settings in cluster locating phase on SIFT100M.}
\label{fig13}
\end{figure}

\textbf{Exploration of the amount of sub-spaces:} The amount of sub-spaces is affected by slices divided in dimension and cluster-based sub-spaces in each vector slice. For the amount of slices in dimension, since vectors have already been partitioned by dimension during PQ encoding in LC, the corresponding vector space is smaller than CL. Therefore, we focus on CL here. As shown in Figure~\ref{fig13}(a), no sub-spaces benefit from low-precision processing without dimension partition, indicating failure of the prediction model due to frequent intersecting among sub-spaces. With the increase of vector slices to 16, the amount of sub-spaces utilizing low-precision distance calculation increases to 87.50\%, with overall accuracy loss kept below 3.98\%. Note that when the amount of vector slices increases to 32, the proportion of low-precision distance calculation decreases, indicating shrunk chances for mixed-precision acceleration with too many vector slices. When the amount of sub-spaces in each vector slice increases from 64 to 512, as shown in Figure~\ref{fig13}(b), the amount of sub-spaces utilizing low-precision distance calculation increases from 75.00\% to 87.50\%, with overall accuracy loss kept below 3.98\%, illustrating more chances for mixed-precision acceleration. LC indicates similar trends, though the vector space is smaller than CL.

\textbf{Kernel function of the regression model:} We utilized linear kernel function for the regression model at first, with the same iteration limits as RBF kernel and default hyperparameters ($C = 0.6, toler = 0.001$). However, no sub-spaces benefit from low-precision processing in this case, indicating non-linear relation between the extracted features and the proper precision to be predicted. On the other hand, as shown in Figure~\ref{fig12}, when equipped with RBF kernel, more than 74.98\% of distance calculation benefits from low-precision processing in either CL or LC, and the proportion would even increase to more than 93.75\% in LC. Thus, we chose RBF kernel function as default for the regression model.

\begin{figure}[!t]
\centering
\includegraphics[width=3.3in]{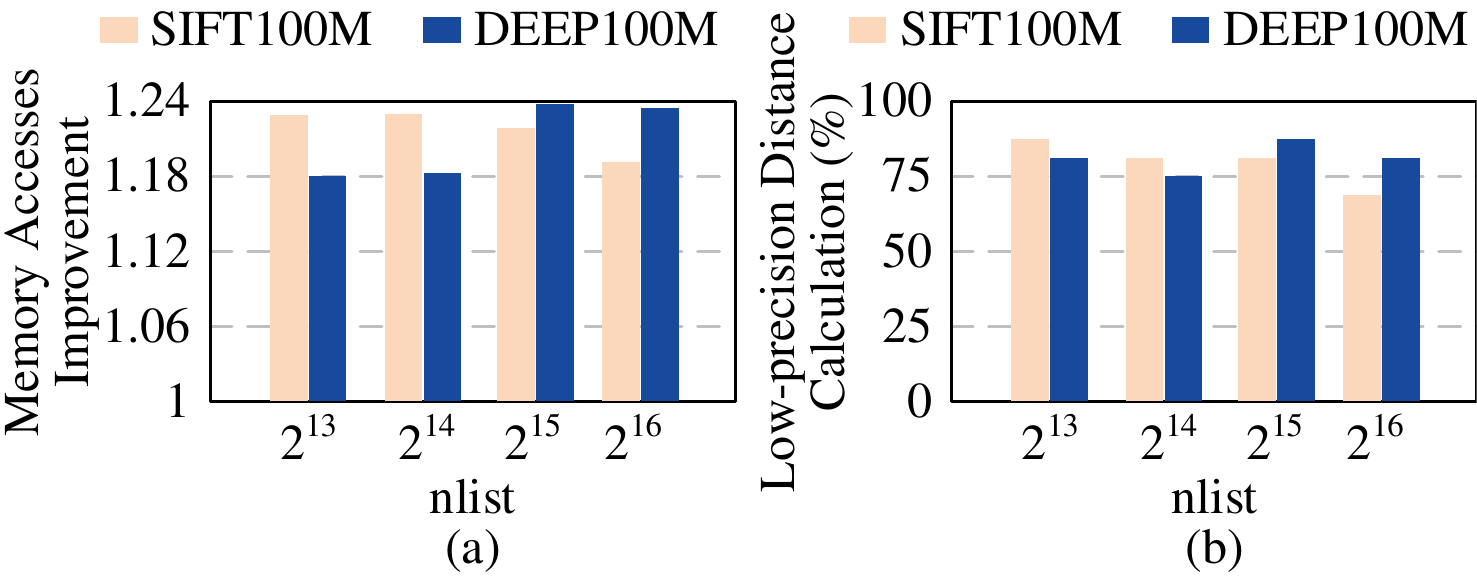}
\caption{\# of memory accesses under different data layouts and datasets.}
\label{fig14}
\end{figure}

\textbf{Data layout optimization:} Vectors using low-precision distance calculation only require a part of bits, leading to discontinuous memory accesses with ordinary data layout that hurts bandwidth utilization, or a waste of bandwidth with loading unnecessary bits for serial memory accesses. In this experiment, we compare the adopted bit-interleaved data layout as mentioned in Section~\ref{sec:hardware:memory} with ordinary data layout. For efficient utilization of bandwidth, we pad loaded bits with ordinary data layout. Therefore, the number of overall memory accesses can be used to measure bandwidth efficiency. Since the loaded data come from on-chip buffer in LC, we focus on benefits in CL without buffer optimization here. As shown in Figure~\ref{fig14}, the adopted bit-interleaved data layout achieves more than 1.18 $\times$ memory efficiency than ordinary data layout on both datasets. Specifically, on SIFT100M, bandwidth efficiency is generally linear to the proportion of sub-spaces utilizing low-precision processing among various algorithm configurations, indicating that the alternatives of precision among all sub-spaces are insensitive to index configurations. On DEEP100M, when $nlist$ is small, sub-spaces utilizing low-precision processing tend to choose conservative bit-width, which affects bandwidth benefits, leading to inconsistency of trends between bandwidth efficiency and the proportion of low-precision sub-spaces. In general, mixed-precision processing improves bandwidth efficiency of ANNS with the adopted bit-interleaved data layout.

\begin{figure}[!t]
\centering
\includegraphics[width=3.3in]{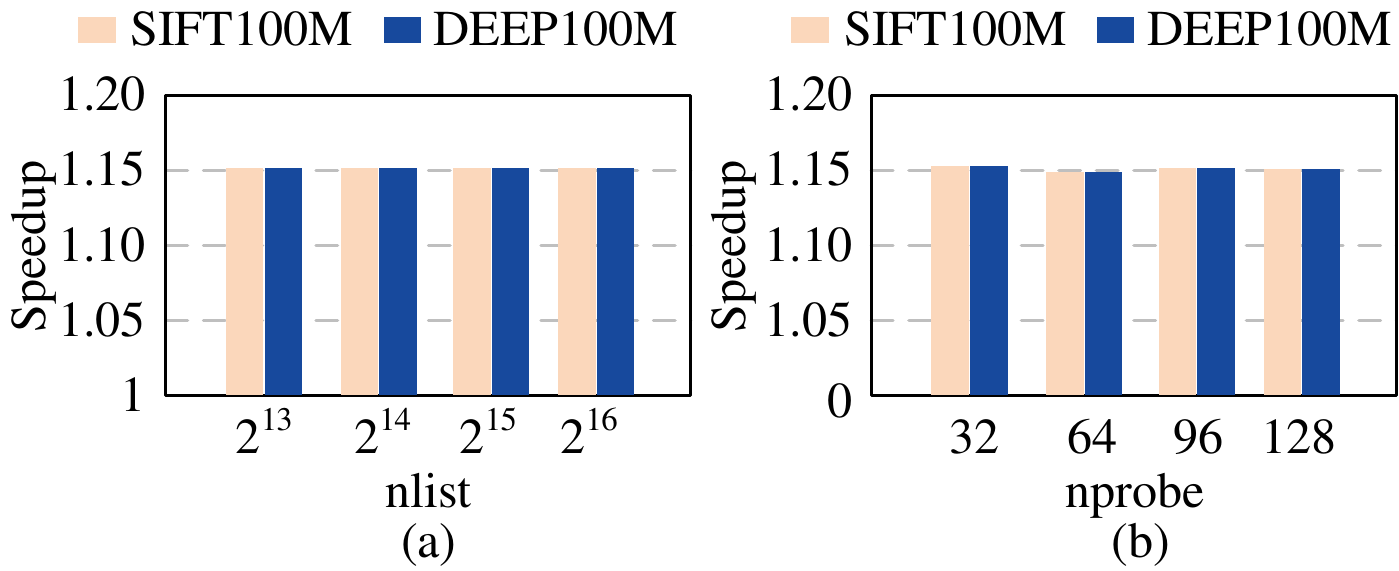}
\caption{Speedup of Load Scheduling Module (LSM) on the look-up table construction phase with loose accuracy constraints.}
\label{fig15}
\end{figure}

\textbf{Load-balance optimization:} As mentioned in Section~\ref{sec:evaluation:setup}, datasets are quantified to uint8, and we adopt a high-quality accuracy constraint as default. Therefore, the choices of sub-spaces for low-precision processing focus on several conservative kinds of bit-width, leading to a balanced load distribution among processing units. As a result, the proposed load balance module has negligible benefits in CL, and improves only 2.76\%-3.56\% of throughput in LC. However, with loose accuracy constraints, sub-spaces have more alternatives of precision, resulting in demands of load balance. As shown in Figure~\ref{fig15}, in this case, LSM improves the throughput in LC by 1.148 $\times$-1.153 $\times$ across various index configurations. However, the benefits on CL are still negligible, indicating that the default order of index vectors in each slice is approximately in coincidence with those reordered by corresponding sub-spaces. In summary, LSM enables ANNS-AMP to be efficiently applied to applications with more precision alternatives, especially when datasets are quantified to more bits (such as uint16) or the accuracy constraints are loosed.

\subsection{Scalability}

\begin{figure}[!t]
\centering
\includegraphics[width=2.8in]{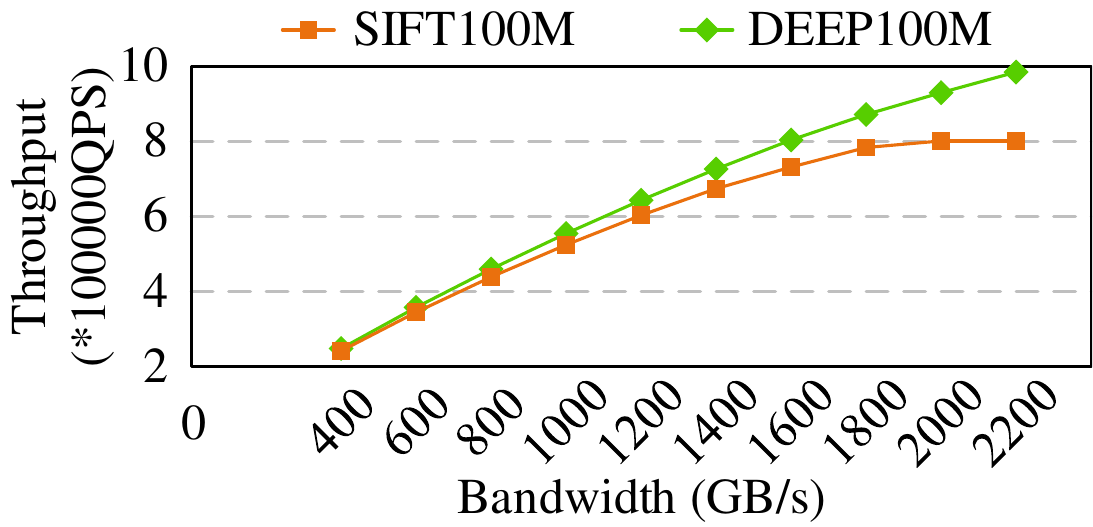}
\caption{Relation between the throughput and bandwidth of ANNS-AMP.}
\label{fig16}
\end{figure}

\textbf{Enlarged bandwidth:} In this experiment, we explore the demanded bandwidth of ANNS-AMP. As shown in Figure~\ref{fig16}, the throughput gets saturated when the bandwidth reaches 2000GB/s with default ANNS configurations. When the bandwidth exceeds 2000GB/s, ANNS-AMP can improve the performance by adding sub-spaces or instances for better computational capability. As mentioned in Section~\ref{sec:evaluation:setup}, the power and area of ANNS-AMP is much smaller than CPU and GPU utilizing the same process technology, leaving chances for extension to memory with higher bandwidth.

\textbf{Comparison with the state-of-the-art accelerator with alternative ANNS index:} To our best knowledge, ANNA \cite{refp10} represents the state-of-the-art (SOTA) accelerator for cluster-based ANNS index. Among recently proposed ANNS accelerators, we further compare ANNS-AMP with Ansmet \cite{refp24}, a graph-based ANNS accelerator, to demonstrate the effectiveness of our design. For a fair comparison, we restrict the bandwidth of ANNS-AMP to match that of Ansmet, and evaluate performance speedup utilizing million-scale datasets, SIFT1M and GIST1M \cite{refp21}, with dimensions of 128 and 960 respectively. The performance of Ansmet on these datasets is estimated from results reported in its original paper. As shown in Table~\ref{table3}, ANNS-AMP achieves a speedup of up to 155.48 times over Ansmet across different datasets and accuracy constraints, owing to its tailored design for cluster-based ANNS index with adaptive mixed-precision computing and memory optimizations. In contrast, Ansmet relies on a graph-based ANNS index, which incurs numerous random memory accesses during vector retrieval and exacerbates the IO bottleneck. In addition, on larger datasets, ANNS-AMP stands to benefit further from large batches of distance calculation and increased opportunities for low-precision arithmetic.

\begin{table}[t!]
  \centering
  \caption{Speedup over Ansmet \cite{refp24} with different recall@10.}
  \label{table3}
  \begin{tabular}{cccc}
    \hline
    recall@10 & 75\% & 80\% & 85\%\\
    \hline
    SIFT1M & 52.86 & 61.68 & 155.48 \\
    GIST1M & 7.33 & 11.16 & 24.80 \\
    \hline
  \end{tabular}
\end{table}

\section{Related Works}

\subsection{Approximate Nearest Neighbor Search}

To make approximate nearest neighbor search (ANNS) efficient, prior works have proposed numerous algorithms, including tree-based \cite{refp14, refp25, refp26, refp27, refp28, refp29}, hash-based \cite{refp30, refp31, refp32}, graph-based \cite{refp33, refp34, refp35, refp36, refp37, refp38}, and cluster-based \cite{refp39, refp40, refp41}. Among them, recent works have proved that cluster-based index is efficient for ANNS on large-scale high-dimensional vector corpus with the same memory footprints \cite{refp42, refp10, refp11}. Despite the efficiency, cluster-based index suffers from redundant distance calculation of non-neighbor vectors, which could take a proportion of more than 90\% on typical datasets \cite{refp24}. Therefore, some works \cite{refp10, refp04, refp06, refp02, refp09} attempted to reduce these redundant search by tuning index parameters. Among them, ANNA \cite{refp10} and Faiss \cite{refp04} enable various combination of index parameters. FANNS \cite{refp06} explores legal index parameters exhaustively with the accuracy constraints based on a sample query set and generates the optimal accelerator design through the predicted throughput and resource consumption with a series of basic hardware blocks. VDTuner \cite{refp02} sets both speed and accuracy of ANNS as objectives and explores the optimal combination of index parameters on CPU by Bayesian optimization. In each iteration of the exploration, VDTuner builds an ANNS index with candidate index parameters and evaluates the speed and accuracy by execution with a typical ANNS library. DRIM-ANN \cite{refp09} evaluates search speed by an analytical performance model to avoid expensive index building and search on physical machine and explores the optimal index parameters by Bayesian optimization as well. However, index parameters are fixed during ANNS, limiting their abilities to dynamic variation of query distribution. Thus, several works apply early termination to ANNS based on runtime search. For instance, VBASE \cite{refp01} maintains a traversal window of distance between the query and latest vectors, and terminates when the median distance in the window become stable. Adaptive Beam Search \cite{refp03} terminates search when current distance of the candidate vector from the query exceeds $(1 + \gamma)$ times than the maximal distance maintained in the candidate neighbor list, and obtained the upper bound of $\gamma$ via theoretical analysis. These works focus on overall algorithmic pruning and depend on a top-\emph{k} list for pruning, limiting their ability to reduce redundancy in ANNS phases without top-\emph{k} lists. Compared with them, ANNS-AMP reduces redundancy by adapting the appropriate precision to distance calculation based on a prediction model that is independent on top-\emph{k} lists at runtime, enabling redundancy reduction of distance calculation in ANNS phases either with or without a top-\emph{k} list.

JUNO \cite{refp05} notices the redundancy in a specific phase of ANNS on the GPU RT core, LUT construction. It divides codebook vectors into 2-dim slices and partitions them by grids. Based on the distance from the grid that the query falls in, JUNO prunes LUT construction of far grids through the distance threshold predicted by a linear regression model. However, high-dimensional neighbors are not always near to queries at each 2-dimensional grid-based space, and pruning across these grids makes the LUT incomplete to calculate high-dimensional distance between them, limiting its ability to reduce the redundancy in distance calculation. Compared to it, ANNS-AMP adopts cluster-based space partition for effective feature extraction and precision prediction to detect the redundancy of distance calculation, and keeps the complete LUT for high-dimensional distance accumulation with redundancy reduction via adaptive low-dimension calculation. 

\subsection{Bit-serial Computation}

Bit-serial arithmetic provides accesses to dynamic trade-off between search speed and accuracy at the circuit level by controlling bit-width for distance calculation. Since the bit-width is determined by algorithmic level, bit-serial computation bridges software ANNS algorithm and hardware processing units for fine-grained acceleration. Prior works have applied bit-serial computation to low-dimensional ANNS. Among them, BitNN \cite{refp43} utilizes the most significant bits (MSBs) of the query and vectors in the point cloud to estimate the lower bound of Euclidean distance between them, and terminates distance calculation and bit loading once the partial distance exceeds the maximal of candidate top-\emph{k} nearest neighbors. It also applied a bit-interleaved mapping on SRAM to avoid bit-padding. PICK \cite{refp44} proposes a bit-serial-based bit-width clipping algorithm with OR operations to clip 16-bit operands to a target bit-width, which is obtained by experiments with the accuracy constraints. It equips a large on-chips SRAM to avoid off-chip memory accesses and applies a vertically aligned bit layout for efficient bit-width clipping. However, ANNS on high-dimensional vector corpus has different features from low-dimensional data, leading to high computational complexity and capacity consumption with naive transfer of low-dimensional methods. In high-dimensional scenes, Ansmet \cite{refp24} eliminates common prefix among vectors offline. At runtime, it fetches MSBs of remaining part with a coarse-grained step since they are expected to have high entropy, and it fetches the least significant bits (LSBs) with a fine-grained step to terminate loading and computing timely. Compared to these works, ANNS-AMP predicts the appropriate bit-width of index data at the granularity of sub-spaces, enabling efficient batch processing and serial memory accesses for low-precision vectors. Meanwhile, the bit-serial early termination technologies in prior works also depend on top-\emph{k} lists to determine the termination condition, limiting their use across multiple ANNS phases. 

\section{Conclusion}

In this paper, we propose ANNS-AMP, an ANNS accelerator in adaptive mixed-precision. It predicts the appropriate bit width by a regression model based on the distance between the query and index vector in each phase of ANNS according to representative features of cluster-based vector sub-spaces. At the architecture level, we propose a specific design utilizing bit-serial computational units with a bit-interleaved layout for acceleration of distance calculation in low-precision sub-spaces, and alleviate load imbalance caused by variation of bit-width among computational units with greedy schduling. Experiments show that ANNS-AMP achieves 163.76 $\times$, 10.57 $\times$, and 2.06 $\times$ performance speedup and 1100.00 $\times$, 39.41 $\times$, and 6.66 $\times$ energy efficiency improvement on average over CPU, GPU, and ASIC, respectively. Specifically, the proposed precision prediction strategies benefits ANNS-AMP from low-precision on up to 87.49\% and 93.75\% of distance calculation in CL and LC phases of ANNS respectively with the overall accuracy loss kept below 2.7\%.


\bibliographystyle{ACM-Reference-Format}
\bibliography{ANNS_AMP}

\end{document}